\input{aipcheck}

\documentclass[
    ,draft            % use draft while you are working on the paper
  ]
  {aipproc}

\layoutstyle{6x9}

\usepackage{amsmath,amssymb}

\begin{document}

\title{Black hole/string ball production, \\ possibly at LHC}


\author{Kin-ya Oda}{
  address={Department of Physics, Osaka University, Osaka 560-0043, Japan}
}

\begin{abstract}
I show a brief historical overview of recent developments on the black hole physics that can be possibly explored at LHC. I comment on the correspondence principle of black holes and strings and show its realization in a differential production cross section of a black hole/string ball with fixed angular momentum.
\end{abstract}

\maketitle

%%%%%%%%%%%%%%%%%%%%%%%%%%%%%%%%%%%%%%%%%%%%
%% MAINMATTER
%%%%%%%%%%%%%%%%%%%%%%%%%%%%%%%%%%%%%%%%%%%%

\section{Introduction}
It is one of the most profound questions how to realize a fully quantum description of gravity.
Superstring theory is a milestone in this path, which can compute scattering of gravitons on a fixed spacetime background, canceling all the possible divergences up to a few loops, see e.g.~\cite{Adam:1900zz} and references therein.
It is even conjectured that a full quantum gravity, including stringy excited modes, is dual to a boundary quantum field theory through the Anti de Sitter space/Conformal Field Theory (AdS/CFT) correspondence~\cite{Maldacena:1997re}.
However it is yet unclear how one can fully treat the dynamics of the background spacetime particularly including a process that changes the boundary.
The difficulty seems to reside in handling the truly non-perturbative nature of gravitation which would play essential role in, e.g., the formation of a black hole.
There are proposals on the non-perturbative formulation of superstring theory~\cite{Banks:1996vh,Ishibashi:1996xs} but no one has succeeded to show that such a model can really give all the results of a perturbative superstring theory in a self-consistent manner.
Given the decades of theoretical struggle and inaccomplishment after the claim that the superstring theory is the theory of everything, it is definitely wanted to have \emph{experimental} data that give a clue to the non-perturbative dynamics of the quantum gravity.
Black hole production from a collision of elementary particles, described by quantum fields, will be such a process if it is realized at the ongoing CERN Large Hadron Collider (LHC).

In superstring theory, there are ten spacetime dimensions.
If some of these dimensions are compactified with large radius~\cite{Antoniadis:1998ig} and/or warped geometry~\cite{Randall:1999ee}, higher dimensional Planck scale may be around TeV for the Standard Model (SM) fields localized at a brane and may become accessible for the LHC experiment.
When the center of mass energy of a collision exceeds the Planck scale, resultant black hole will have a horizon radius larger than its Compton wavelength and therefore quantum fluctuation cannot prevent the BH to form.
Above the Planck scale, the black hole production cross section would be well-treated by the classical production cross section that \emph{grows} with center of mass energy.
The higher the center of mass energy above the Planck scale, the larger the produced black hole becomes and hence the better be its classical treatment.
Well above the Planck scale, this growing classical cross section is proven in four dimensions~\cite{Eardley:2002re} and in higher dimensions~\cite{Yoshino:2002tx}, within the assumption that the initial particles entailing gravitational shockwaves can be treated by the Aichelburg-Sexl solutions; See also~\cite{Giddings:2004xy} for a justification utilizing quantum wave packets for initial particles and~\cite{Giddings:2009gj} for a gravitational S-matrix.

\section{Predictions of Black Hole Events at LHC}
Let us present a brief historical overview of recent developments on the black hole physics at LHC.
The theoretical importance of the black hole production process is first advocated by 't~Hooft~\cite{'tHooft:1987rb} and is revisited in the context of TeV scale gravity scenario in~\cite{Banks:1999gd}, where it was considered that the TeV BH would radiate mainly into bulk graviton modes.
See also~\cite{Argyres:1998qn} for a study of the TeV BH phenomenology under this assumption.
Lately it is found that such a BH radiates mainly on the Standard Model (SM) fields that are located on the brane~\cite{Emparan:2000rs}.
Following this observation, theoretical aspects of the BH production event is considered in~\cite{Giddings:2001bu}, while the possible LHC signals are studied in~\cite{Dimopoulos:2001hw} in the approximation that BH decays instantaneously with a blackbody spectrum.\footnote{
In~\cite{Oda:2001uw} we have studied the impact of the stringy suppression of the hard scattering processes and considered possible link to the BH production process.
}

The true interest is the non-perturbative aspects of the quantum gravity, which in real experiment can only be revealed as discrepancies from the semi-classical prediction of the BH production and decay, given the current lack of theoretical knowledge in this non-perturbative region around the Planck scale.
(Namely, in this prediction we treat the produced black hole as a classical background while the SM matters as quantum fields on the fixed curved background, leaving all the gravitational dynamics left behind.)
For this purpose of extracting discrepancies, it is important to predict the BH event within the semi-classical Hawking radiation picture as precisely as possible.
In the approximation that BHs radiate mainly on brane~\cite{Emparan:2000rs}, the Hawking spectrum is completely fixed when one computes the greybody factors for the fields located on the brane.
In~\cite{Kanti:2002nr}, the authors have obtained the field equation and the resultant greybody factors for a brane scalar that should correspond to the SM Higgs field.
In~\cite{Ida:2002ez}, we have obtained the separable master equation for brane-localized spinor and vector fields that correspond to the main BH decay modes: the SM quarks, leptons, and gauge bosons.\footnote{
In~\cite{Ida:2002ez} there was a typo in the master equation~\cite{PhysRevD.69.049901}, while all the calculations, results and conclusions are unchanged. The typo correction~\cite{PhysRevD.69.049901} seems to have been overlooked in Eq.~(18) in~\cite{Casals:2005sa}. The result for both helicities is summarized in~\cite{Ida:2006tf}.
}
We have argued that the black holes tend to be produced with large angular momenta and then have shown that the resultant Hawking spectra are highly anisotropic, within a low frequency approximation.\footnote{
In~\cite{Ida:2002ez} it has been noticed that ``When we average over opposite helicity states, the up-down asymmetry with respect to the angular momentum axis ... disappears although there still remains the angular dependence itself,'' while this statement seems to be overlooked in~\cite{Casals:2005sa}.
}
In~\cite{Ida:2006tf}, we have spelled out the full Hawking spectra, except for the bulk graviton emission, by solving the radial part of the master equation numerically.
Our claim that the Hawking radiation has strong anisotropy, based on the low frequency expansion, is confirmed by numerical computations on the angular equation for vector~\cite{Casals:2005sa}, while it becomes milder for spinor~\cite{Casals:2006xp}.
Our result is implemented in the black hole event generator BlackMax~\cite{Dai:2007ki}, which now ``incorporates the effects of black-hole rotation, splitting between the fermions, non-zero brane tension and black-hole recoil due to Hawking radiation (although not all simultaneously).''

\section{Correspondence to String Ball Production?}
Black hole production process is theoretically important in the sense that it is dual to a production of a heavy string state, given the correspondence principle of black holes and strings, whose importance is first advocated by 't Hooft~\cite{'tHooft:1990fr}.
At low energy $E\ll M_s/g_s^2$, the perturbative string theory would give a good description of nature while at high energy $E\gg M_s/g_s^2$, particle scattering would be well approximated by a non-perturbative but classical gravitational black hole formation process, where $M_s\equiv(\alpha')^{-1/2}$ is the fundamental string scale and $g_s$ is the (closed) string coupling constant.
In the intermediate region $E\sim M_s/g_s^2$ both the string and black hole pictures are broken down but still extrapolations from these two pictures give the same order of physical quantities such as temperature, entropy, size, production cross section, etc., see e.g.~\cite{Matsuo:2008fj} for review and references.\footnote{
The final state energies from a string scattering obey a sort of anti-scaling low when one increases the center of mass energy in the region $M_s/g_s\lesssim E\lesssim M_s/g_s^2$~\cite{Veneziano:2004er}.
This might indicate another support for the correspondence in the sense that the higher energy implies the larger mass for the corresponding BH and hence the lower Hawking temperature. It is suggestive that the scale $M_s/g_s$ is the scale above which D-brane interactions become significant.
}

In the following I review our recent study on the correspondence in the differential production cross section with a fixed angular momentum~\cite{Matsuo:2008fj}.
When both the initial particles and the final BH can move in $D$ spacetime dimensions, we get the following geometrical cross section for production of a BH with an angular momentum $J$ with the center of mass energy squared $s$~\cite{Ida:2002ez,Matsuo:2008fj}:
\begin{align}
	{d\sigma_\text{BH}\over dJ}\sim {J^{D-3}\over s^{(D-2)/2}},
  \end{align}
when $J$ is enough smaller than the maximum value $J_\text{max}$ shown in~\cite{Ida:2002ez}.
On the other hand, it is well known that a string theory amplitude gives, through the optical theorem, the following exponentially soft behavior:
\begin{align}
	\sigma_J(s)\propto e^{-J^2/s\ln s}.
  \end{align}
Here and hereafter, we drop all the irrelevant numerical factors for qualitative argument such as 2, $\pi$, etc.\ and take the stringy natural unit $M_s=1$.
The BH geometrical cross section increases with $J$ while the stringy cross section decreases exponentially.

To investigate this apparent contradiction, we have performed the application of the optical theorem to the closed string two-to-two amplitude, after expanding it into partial waves with angular momentum $J$. The result we obtain for $J\gg 1$ and $s\gg M_s^2$ is
\begin{align}
	\sigma_J(s)\sim J^{D-3}{g_s^2s\over (s\ln s)^{D-2\over2}}e^{-J^2/s\ln s},
  \end{align}
where the prefactor of the exponential is what we newly find.
We can see that in the region $1\ll J\ll s\ln s$, where the exponential softness for large $J$ is not yet dominant, we obtain the geometrical behavior of the BH differential cross section $\propto J^{D-3}$, as asserted in~\cite{Ida:2002ez}, from completely different stringy point of view.\footnote{
Even neglecting the $\ln s$ correction, the $s$ dependence shows that the absolute values of the differential cross section in the black hole and string pictures match only at $\sqrt{s}\sim M_s/g_s$, rather than at the correspondence scale $M_s/g_s^2$. Its physical interpretation is yet unclear.
}

If we take the correspondence literally, the black hole production will be dual to a string ball production at the correspondence scale.
In string amplitude, production of a heavy (long) $s$-channel resonance is dual to a $t$-channel exchange of a single massless low energy graviton.
It is shown that resummation over infinite number of $t$-channel graviton exchange gives a propagation under black hole background, see~\cite{Amati:2007ak} and references therein.
It is still to be clarified how these two pictures, i.e.\ an exchange of single long-stretched string vs. infinite number of exchanges, can match each other.

\section{Summary and Outlook}
It is possible that the higher dimensional Planck scale is around TeV which is directly accessible for the LHC.
Well above the Planck scale, particle scattering will be dominated by the black hole production and its subsequent decay via the Hawking radiation.
The black hole production well above the Planck scale can be treated by a classical but non-perturbative Einstein gravity.
Once produced, a black hole decay can be treated by the Hawking radiation, which is now completed by ourselves and by colleagues and is now implemented in the event generators.

I note that the real data from LHC, if any, will not precisely follow the above black hole prediction since its semi-classical treatment is marginal at best.
This can be seen e.g.\ from the fact~\cite{Ida:2002ez} that the largest possible angular momentum $J_\text{max}$, corresponding to the maximum impact parameter, is limited by $J_\text{max}\lesssim 10$ in units of $\hbar$ at the LHC energy, which is by no means classical.\footnote{
The limitation of the semi-classical picture is emphasized in~\cite{Meade:2007sz}.
}
As is emphasized in~\cite{Ida:2002ez}, our true interest is a deviation from the semi-classical black hole picture that would give a clue to the non-perturbative dynamics of quantum gravity at the so-called Planck phase.
For that purpose, we have fixed the Hawking radiation from a TeV black hole.\footnote{
There was claim that the black hole decay spectrum is smeared by a quark-gluon plasma called the chromosphere. In~\cite{Alig:2006up} we have shown that indeed there will occur large number of interactions among hard quarks/gluons but the jet structure is still preserved and that the chromosphere does not form after LHC black hole.
}

One of the remaining tasks is to predict the non-perturbative formation process of a black hole.\footnote{
There also remains to compute the Hawking radiation spectrum for the bulk graviton emission, which will be greatly enhanced when black hole is highly rotating.
}
This process will be well-treated by the classical Einstein gravity in higher dimensions if the scattering is much above the higher dimensional Planck scale.
See~\cite{Shibata:2008rq} for recent development of the numerical simulation.

As said above, the actual process at LHC will involve quantum corrections to the classical result.
Correspondence to the production of a heavy string state might gives some hint.
I have briefly reviewed our recent result on the geometrical behavior of the black hole/string ball production cross sections.
It would be great if the LHC finds BHs and even if not, the BH formation process will continue to be one of the most important theoretical ground for achieving the goal of quantum gravity.

\bibliography{odakin}
\bibliographystyle{aipproc}

\end{document}